\documentclass[twocolumn,showpacs,preprintnumbers,amsmath,amssymb,pra]{revtex4}

\usepackage{graphicx}
\usepackage{dcolumn}
\usepackage{bm}
\usepackage{changebar}

\usepackage{amssymb,amsmath}
\usepackage{wasysym}
\usepackage{verbatim}
\usepackage{threeparttable}

\newcommand{\figwidth}{9.0cm}   

\newcommand{\cm}{cm$^{-1}\ $}

\def\be{\begin{equation}}
\def\ee{\end{equation}}

\def\4he{$^4$He}
\def\3he{$^3$He}
\def\cm3{cm$^{-3}$}

\begin{document}
\title{Charge transfer in cold Yb$^+$ + Rb collisions}

\author{Elvira R. Sayfutyarova}\altaffiliation{Present address: Department of Chemistry, Frick Laboratory, Princeton University, NJ 08544, USA}
\affiliation{Department of Chemistry, M.V.~Lomonosov Moscow State University, Moscow 119991, Russia}
\author{Alexei~A.~Buchachenko}\email[]{alexei@classic.chem.msu.su}
\affiliation{Department of Chemistry, M.V.~Lomonosov Moscow State University, Moscow 119991, Russia} 
\affiliation{Institute of Problems of Chemical Physics RAS, Chernogolovka, Moscow District 142432, Russia}
\author{Svetlana~A.~Yakovleva}
\author{Andrey~K.~Belyaev}\email[]{belyaev@herzen.spb.ru}
\affiliation{Department of Theoretical Physics and Center for Advanced Study, Herzen University, St.~Petersburg 191186, Russia}

\date{\today}

\begin{abstract}
Charge-transfer cold Yb$^+$ + Rb collision dynamics is investigated theoretically using
high-level {\it ab initio} potential energy curves, dipole moment functions and nonadiabatic coupling
matrix elements. Within the scalar-relativistic approximation, the radiative transitions from the entrance
$A^1\Sigma^+$ to the ground $X^1\Sigma^+$ state are found to be the only efficient charge-transfer pathway.
The spin-orbit coupling does not open other efficient pathways, but alters the potential energy curves and the 
transition dipole moment for the $A-X$ pair of states. The radiative, as well as the nonradiative, charge-transfer cross sections calculated within
the $10^{-3}-10$ cm$^{-1}$ collision energy range exhibit all features of the Langevin ion-atom collision regime,
including a rich structure associated with centrifugal barrier tunneling (orbiting) resonances. Theoretical 
rate coefficients for two Yb isotopes agree well with those measured by immersing Yb$^+$ ion
in an ultracold Rb ensemble in a hybrid trap. Possible origins of discrepancy in the product distributions
and relations to previously studied similar processes are discussed. 
\end{abstract}

\pacs{34.20.Cf,34.70.+e,34.50.Cx}
\maketitle

\section{Introduction}

Ongoing efforts in the creation of cold atomic and ionic ensembles, as well
as in their subtle manipulation, have been recently merged in preparation and
study of the hybrid ion-atom systems
\cite{Willitsch2008,Grier2009,Zipkes2010Nature,Zipkes2010PRL,Schmid2010,Rellergert2011,Hall2011,Ratschbacher2012,Ravi2012, Hall2013arXiv}.
This signifies the important step towards probing, understanding and exploring
ion-neutral interactions and collisions at sub-Kelvin temperatures. Elastic
and momentum transfer collisions may serve as the means of cooling of both
neutral
\cite{Moriwaki1992,Molhave2000,Makarov2003,Smith2005}
and charged
\cite{Zipkes2010Nature,Zipkes2010PRL,Ravi2012}
components. Ion-atom collisions resulting in charge transfer are of interest for a prototype of
reactive processes and suggested to lead to gas-phase ``metal-to-insulator''
transitions at microKelvin temperatures
\cite{Cote2000}.

Elastic, as well as resonant charge transfer, ion-atom collision dynamics is well understood theoretically
\cite{Makarov2003,CoteDalgarno2000}.
High-energy limit mediated by a short-range exchange interaction transforms
to the so-called Langevin regime when a collision energy becomes comparable to 
a long-range induction interaction. When a collision energy decreases further,
the quantum regime with its threshold laws for each partial wave is attained.
This limit has been thoroughly considered within the multichannel quantum defect theory
\cite{Idziaszek2009,Gao2010,Idziaszek2011,LiGao2012},
but not yet reached experimentally.

Preparation of cold ion-atom systems is achieved in a hybrid trap by bringing in contact atomic and ionic ensembles or by creating ions via photoionization of cold atoms
\cite{Grier2009,Zipkes2010PRL,Schmid2010,Hall2011,Makarov2003,Smith2005,Sullivan2011}.
Collision- or photo-induced processes in Coulomb crystals
can also be studied below 10 K \cite{Willitsch2008,Staanum2008,Willitsch2008PCCP}
providing that the neutrals are cooled or decelerated
\cite{Hall2012}. 
However, in all cases the motion of ion(s) in the trapping field creates an unavoidable inherent source of a kinetic energy that imposes the limits on cooling
\cite{Cetina2012}. 
Effective temperatures of a few tens of milliKelvin attained experimentally
\cite{Grier2009,Schmid2010}
still correspond to the Langevin regime with its complex dynamics sensitive
to the global ion-atom potentials, transition moments and coupling matrix
elements. In addition, laser cooling creates appreciable amount of electronically excited species, which further complicate the dynamics
\cite{Hall2011,Sullivan2012}.

This explains the need in thorough theoretical studies that account for specific features
of the particular perspective system. Resonant
charge transfer has received more attention, with the focus on the model
alkali or alkaline-earth (e.g.,
Refs.~\cite{CoteDalgarno2000,Zhang2009JPCA,Zhang2011})
and Yb$^+$ + Yb
\cite{Zhang2009}
systems. For nonresonant case, few combinations of alkaline-earth ions and
alkali atoms 
\cite{Makarov2003,Smith2005,Belyaev2011,Krych2011,Rakshit2011,Belyaev2012},
as well as Yb$^+$, Ba$^+$ + Ca collisions
\cite{Rellergert2011,Sullivan2012},
were considered.
Accurate {\it ab initio} calculations that back most of these studies revealed
the complexity of charge-transfer pathways in the above systems. Strongly
bound excited electronic states may well be involved in the dynamics by the
nonadiabatic and/or spin-orbit couplings, or even directly upon laser cooling, preventing the use of simple
two-state models
\cite{Rellergert2011,Sullivan2012,Belyaev2011,Krych2011,Belyaev2012}.
Collision dynamics is complex itself being dominated by the resonances typical
to the multiple-partial wave Langevin regime. 

In the present paper we expand the theoretical experience considering the charge transfer in cold
collisions between the ground-state Yb$^+$ ion and Rb atom. Experimental studies by
K\"ohl and coworkers 
\cite{Zipkes2010Nature,Zipkes2010PRL}, 
further extended to excited Yb$^+$ ions
\cite{Ratschbacher2012},
demonstrated the possibility of cooling an Yb$^+$ ion in the ultracold Rb
environment, gave the charge-transfer rate coefficient of the order of
10$^{-14}$ cm$^3$/s (the smallest one measured so far below 1 K) and
characterized the product distributions. This system is attractive for
further experimental studies since both Yb($^1$S) and Rb($^2$S) neutrals
can be brought to degenerate gases
\cite{Takasu2003PRL,Fukuhara2007PRL,Anderson1995Science}
and combined with Rb$^+$($^1$S) and Yb$^+$($^2$S) ions for studying
charge transfer from both ``sides''. Rich isotope variety enables one to
address the isotope and hyperfine structure effects.
We performed accurate {\it ab initio} calculations of the potential energy curves,
dipole moments and nonadiabatic coupling matrix elements for the lowest electronic states
of the  (YbRb)$^+$ ion. We showed that within the framework of the scalar-relativistic approximation
the charge transfer can be considered by means of the two-channel model and that
its radiative pathway is barely dominant. The vectorial spin-orbit interaction affects
the electronic structure of the (YbRb)$^+$ ion, but does not produce a remarkable
effect on the low-energy radiative charge transfer. Quantum scattering 
calculations within the range of collision energies relevant to the experiment show a rich resonance structure associated with orbiting
resonances typical for the Langevin regime.  Calculated rates
agree well with the measured ones
\cite{Zipkes2010PRL},
though reveal less pronounced isotope effect. 
On the other hand, our study of single-collision dynamics results in some deviation in interpretation of the product distributions deduced experimentally.

\section{ELECTRONIC STRUCTURE}
The electronic structure of the (YbRb)$^+$ ion was investigated {\it ab initio} using the MOLPRO program package 
\cite{MOLPRO}.
All calculations were performed within the $C_{2v}$ symmetry group with the origin of 
the electronic coordinates placed at the center of nuclear mass of the system composed of $^{174}$Yb and $^{85}$Rb isotopes 
and $z$ axis oriented along the internuclear vector {\bf R}. 

\subsection{Techniques}
Scalar-relativistic (SR) calculations were performed using the small-core 28-electron relativistic core potentials ECP28MDF with the supplementary contracted basis sets for both Yb 
\cite{Wang}
and Rb
\cite{RbBas}. 
The sets of diffuse primitives of each of the $spdf$ types with the exponents continuing two lowest exponents of the standard bases as an even-tempered sequence were added at each center to improve the description of induction and dispersion interactions. 
To evaluate the potential energy curves (PECs) of the states that are the lowest ones in their symmetry (spatial and spin) representation, the restricted version of the coupled cluster method with singles, doubles and noniterative triples, CCSD(T), was employed with the restricted Hartree-Fock (RHF) reference. This type of the calculations always included the set of $3s3p2d2f1g$
bond functions (bf) 
\cite{bf}
placed at the middle of the internuclear distance $R$ and counterpoise correction for the basis set superposition error 
\cite{BoysB}. The Yb($4s^24p^64d^{10}$) shells were included in the core, whereas all the rest electrons were correlated 
explicitly. The lowest excited singlet states were calculated by means of the equation-of-motion approach in coupled clusters with
singles and doubles (EOM-CCSD) implemented in the same way. For excited triplets the multireference configuration interaction
(MRCI) calculations with the Davidson correction 
\cite{Davidson}
were performed with the reference wave functions built by the state-averaged complete active space multiconfigurational 
self-consistent field (CASSCF) method with active space spanned by the Yb($6s6p$) and Rb($5s$) atomic orbitals. The core set in the MRCI calculations consisted of Yb($4s^24p^64d^{10}5s^25p^6$) and Rb(4$s^2$) atomic orbitals, whereas the 4$f^{14}$ shell of an Yb atom was correlated as fully occupied. 

The similar MRCI method was used for spin-orbit (SO) calculations, but with the ECP28MWB effective core potential for Yb containing
the SO part
\cite{ECP28MWB} 
and supplementary segmented basis set 
\cite{MWBSEG}
augmented by the $s2pdfg$ diffuse functions \cite{StructC}. 

The calculations were performed on the fine grid of the internuclear separation $R$ from 1.9 to 40 {\AA}. 

\subsection{Scalar-relativistic results}

The results of SR CCSD(T), EOM-CCSD and MRCI calculations are shown in Fig.~1(a). At low collision energies it is sufficient to consider the states corresponding to the three lowest dissociation limits: 
(i) Yb($^1$S) + Rb$^+$($^1$S) that represents the final charge transfer (CT) channel, 
(ii) Yb$^+$($^2$S) + Rb($^2$S), the initial channel, and 
(iii) closed Yb$^*$($^3$P$^\circ$) + Rb$^+$($^1$S) CT channel that lies slightly above the entrance. 
Our best estimations 
for energies  of excited limits 16279 cm$^{-1}$ (EOM-CCSD) and 18646 cm$^{-1}$ (MRCI) correspond well to the centers of the measured 
fine-structure multiplets, 16750 and 18869  cm$^{-1}$, respectively \cite{NIST}. It is evident that direct CT can only occur through $A^1\Sigma^+$ -- $X^1\Sigma^+$ interactions, either nonadiabatic or dipole ones.

\begin{figure}[ht!]
\includegraphics[angle=0, width=\figwidth]{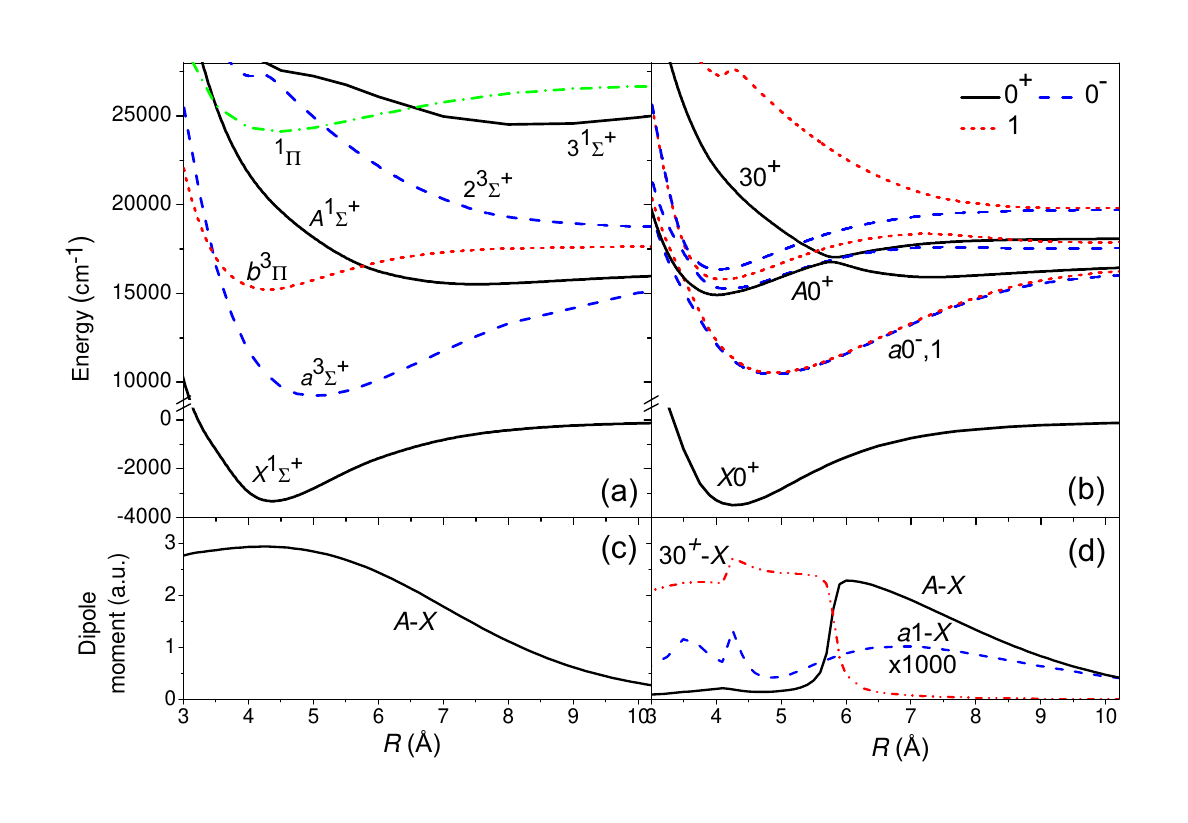}
\caption{\label{fig:all_pots}
(color online) Potential energy curves and dipole moment functions for the lowest electronic states of the (YbRb)$^+$ ion. (a) Scalar-relativistic
PECs. In addition to the states correlating to the three lowest asymptotic limits discussed in the text, $3^1\Sigma^+$ and $^1\Pi$
states correlating to the fourth Yb($^1$P$^\circ$) + Rb$^+$($^1$S) asymptote (26172 cm$^{-1}$, cf. 25068 cm$^{-1}$ 
\cite{NIST}) are shown. (b)  PECs for the lowest SO-coupled 0$^\pm$ and 1 states. (c) The scalar-relativistic $A-X$ transition dipole
moment. (d) The SO-coupled $A-X$, $a1-X$ and $3\,0^+-X$ transition dipole moments. 
}
\end{figure}

To obtain the most accurate SR PECs for the relevant $X^1\Sigma^+$, $a^3\Sigma^+$, $A^1\Sigma^+$, $b^3\Pi$ and 
$2^3\Sigma^+$ states, we took CCSD(T)/bf results for $X$, $a$ and $b$ as the reference. The $A$-state PEC was
obtained by adding EOM-CCSD excitation energies to the ground PEC, whereas the MRCI results for $\Pi-\Sigma$ splitting
were used to obtain $2^3\Sigma^+$ PEC from the CCSD(T)/bf $b^3\Pi$ one. These {\it ad hoc} procedures aim to 
compensate the lower accuracy of EOM-CCSD and MRCI methods and incorporate the basis set superposition error correction. 
The resulting points were 
interpolated by cubic splines and joined smoothly to an analytical long-range (LR) function represented by the lowest-order 
$-C_n/R^n$ asymptotic term. Finally, the PECs were shifted in energy to reproduce the true asymptotic limits. The parameters
of so-obtained PECs are presented in Table \ref{potpar}.

\begin{table}[ht!]
\centering
\caption{
Parameters of the lowest PECs of the (YbRb)$^+$ ion: 
Equilibrium distance, $R_e$, state binding energy, $D_e$, 
leading LR coefficient, $C_n$, energy of the minimum with respect to the ground-state asymptotic
limit, $U_e$, and state dissociation asymptote, $U_\infty$,  with respect to the ground asymptotic limit.
}
\label{potpar}
\begin{threeparttable}
\begin{tabular}{lccccc}
\hline
\hline
State & $R_e$, {\AA} & $D_e$, cm$^{-1}$  & $n$; $C_n$, a.u. & $U_e$, cm$^{-1}$ & $U_\infty$, cm$^{-1}$ \\
\hline
\multicolumn{6}{c}{Scalar-relativistic PECs} \\
\hline
$X^1\Sigma^+$ & 4.28 & 3496 & 4; 72.9 & -3496 & 0 \\
$a^3\Sigma^+$ & 4.89 & 6176 & 4; 163.7 & 10574 & 16750 \\
$A^1\Sigma^+$ & 7.31 & 836 & 4; 155.5 & 15914 & 16750 \\
$b^3\Pi$ & 3.97 & 3218 & 3; -4.6 & 15650 & 18869 \\
$2^3\Sigma^+$\tnote{a} & - & - & 3; 18.8 & - & 18869 \\
\hline
\multicolumn{6}{c}{SO-coupled PECs} \\
\hline
$X0^+$ & 4.28 & 3498 & 4; 72.9 & -3498 & 0 \\
$a0^-$ & 4.87 & 6312 & 4; 163.7 & 10438 & 16750 \\
$a1$ & 4.88 & 6236 & 4; 163.7 & 10513 & 16750 \\
$A0^+$\tnote{b} & 7.30 & 837 & 4; 155.5 & 15912 & 16750 \\
$A0^+$\tnote{c} & 4.04 & 1849 & - & 14901 & 16750 \\
$A0^+$\tnote{d} & 5.78 & - & - & 16790 & 16750 \\
\hline
\hline
\end{tabular}
     \begin{tablenotes}
       \item[a] Repulsive PEC. 
       \item[b] Right minimum, corresponds to the $A^1\Sigma^+$ SR state. 
       \item[c] Left minimum, corresponds to the $b^3\Pi$ SR state. 
       \item[d] Avoided crossing maximum. 
     \end{tablenotes}
  \end{threeparttable}
\end{table}

Accuracy of the {\it ab initio} PECs can be indirectly verified at long distances. LR behavior of the $X$ and $a$, $A$
potentials is dominated by Yb and Rb induction, respectively. The finite-field CCSD(T) calculations estimated the static dipole polarizabilities of neutral atoms as  $\alpha_\mathrm{Yb} = 142.2$ and $\alpha_\mathrm{Rb} = 318.2$ a.u., in perfect agreement with
the recommended values  $\alpha_\mathrm{Yb} = 139 \pm 7$
\cite{Beloy,YbPol}
and  $\alpha_\mathrm{Rb} = 318.8 \pm 1.4$ a.u.
\cite{Holmgren,Deiglmayr}.
The leading induction coefficients $C_4 = \alpha_\mathrm{X}/2$ are equal to 71 and 159 a.u., respectively, in reasonable correspondence
with the fits to {\it ab initio} PECs, see Table \ref{potpar}. Asymptotic dependence of the $b^3\Pi$ and $2^3\Sigma^+$ 
PECs originates from the charge-quadrupole Rb$^+$ + Yb$^*$($^3$P$^\circ$) interaction. The results of Ref.~\cite{YbPol} obtained within the more accurate {\it ab initio}
approach allowed us to estimate the corresponding $C_3$ coefficients as -7.2 a.u. and 14.4 a.u., respectively. They agree with the fitted
coefficients only qualitatively, but the LR behavior of the states correlating to the third dissociation limit is not crucial in the
present context. 

The permanent and transition dipole moments for the pair of $^1\Sigma^+$ states were calculated by EOM-CCSD and MRCI methods, while the finite-field CCSD(T)/bf calculation was possible also for the ground $X$ state. All the methods 
give similar results, though the MRCI $A$-state moment reveals the signatures of the mixing with higher
lying states at the distances shorter than 4 {\AA}. The $A-X$ electronic transition dipole moment $d_{AX}$ responsible
for radiative CT is shown in Fig.~1(c). 

In addition to the transition dipole moment, two $^1\Sigma^+$ states are coupled by the nonadiabatic coupling matrix elements (NACMEs) as defined in Sec.~\ref{sec:nonradiative}. 
The first-order radial NACME was computed using the two-point finite-difference procedure
\cite{MOLPRO}
using the MRCI vectors built on the CASSCF wave functions obtained with averaging over the two lowest $^1\Sigma^+$ states. It was checked that neither inclusion of the third state of the same symmetry into CASSCF averaging nor the use of more accurate three-point differentiation procedure alter the results remarkably. The second-order radial NACME was approximated as the derivative of the first-order one
\cite{Belyaev1999}. 
For two states of the same symmetry the first-order angular NACME, proportional to the orbital electronic angular momentum operators $L_x$ and $L_y$, vanishes by parity. Non-vanishing is the second-order NACME represented by the $L_x^2 + L_y^2$ operator. The one-electron part of the $L_y^2$ operator ($L_x^2$ is the same by symmetry) was calculated as the matrix element on the CASSCF wave functions. As shown in Fig.~2, the first-order radial coupling is expectedly much stronger than the second-order and angular ones. The mixing with the higher lying electronic states at short distances already noted for dipole moments affects NACMEs as well.  

\begin{figure}[ht!]
\includegraphics[angle=0, width=\figwidth]{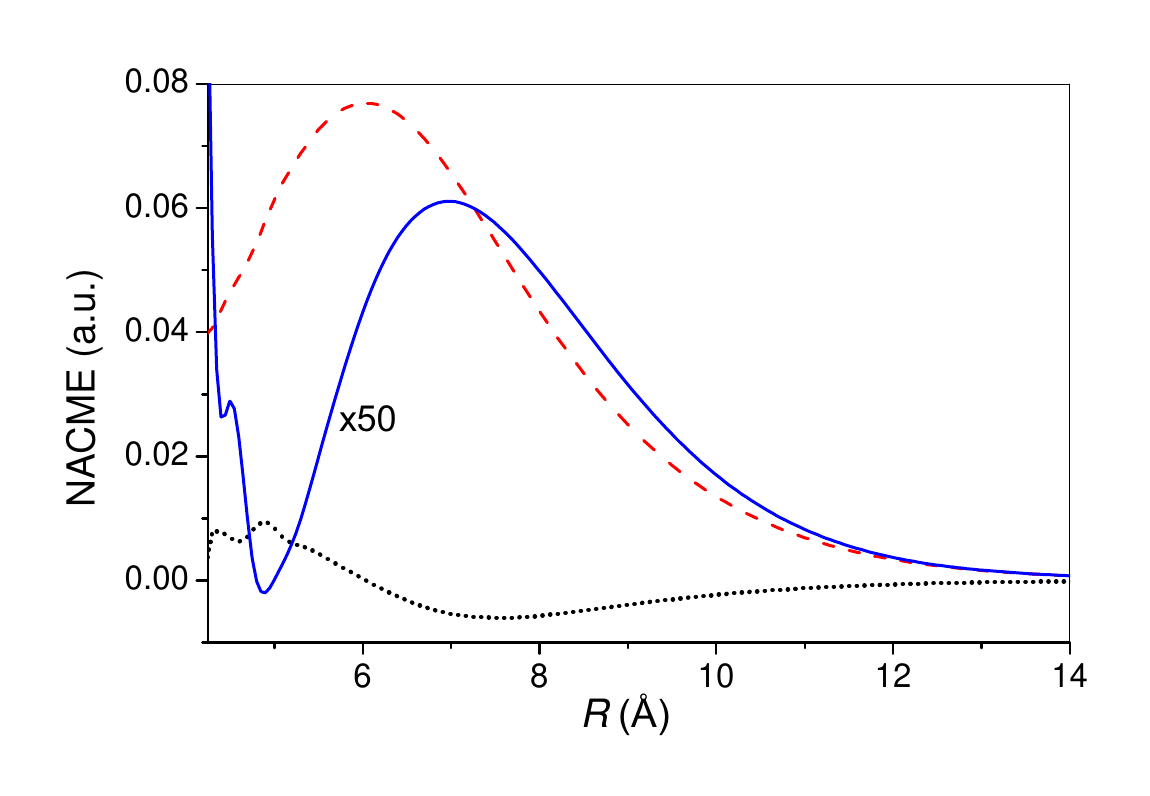}
\caption{\label{fig:nacme}
(color online) Nonadiabatic couplings between the $X^1\Sigma^+$ and $A^1\Sigma^+$ states: the first-order radial NACME (dashed line), the second-order radial one (dotted line) and the second-order angular NACME multiplied by 50 (solid line).
}
\end{figure}

\subsection{Spin-orbit coupling}

The state-interacting MRCI SO calculations 
\cite{SOCI}
were performed in the space spanned by the scalar-relativistic states correlating to three lowest dissociation limits. The SO matrix elements were evaluated for Breit-Pauli operator using the inner part of the MRCI wave functions. The PECs described
above were used as the diagonal SR part of the full Hamiltonian matrix. Relevant SO-coupled PECs are shown in Fig.~1(b) and 
characterized in Table \ref{potpar}.

The states correlating to the lowest dissociation limits are weakly affected by the SO coupling. The only qualitative effect, the
splitting of the $a^3\Sigma^+$ state into the $a0^-$ and $a1$ components, is not significant. Much more pronounced is the 
interaction with the higher states correlating to an excited Yb$^*$ atom. The asymptotic limit splits into three components that
correspond to the $^3$P$^\circ_j$, $j = 0, 1$ and 2, Yb$^*$ fine-structure levels. The lowest $^3$P$^\circ_0$ one gives the single $0^-$
component that may interact with the $a0^-$ state at short distances, whereas the second $^3$P$^\circ_1$ gives $0^+$ and
1 components. The former transforms $b^3\Pi-A^1\Sigma^+$ crossing into the avoided one giving the double-well 
$A0^+$ PEC. The latter perturbs the repulsive wall  of the $a1$ potential. These changes also affect the transition dipole
moments. The $A-X$ transition moment is similar to its scalar-relativistic precursor at distances larger than the avoided crossing
point $R \approx 6$ {\AA} and rapidly falls down at shorter distances, see Fig.~1(d). Instead, a significant dipole coupling 
occurs with the upper $3\,0^+$ state, which acquires charge-transfer character as the result of the crossing. The sum of
$A-X$ and $3\,0^+-X$ transition moments is therefore similar to the scalar-relativistic  $A^1\Sigma^+-X^1\Sigma^+$ one.  
The SO interaction also allows the transition from the $a1$ 
to the $X$ state, but the corresponding dipole moment is negligible. 

\section{Nonradiative charge transfer} \label{sec:nonradiative}

The nonradiative charge transfer process in low-energy Yb$^+$ + Rb collisions can occur due to nonadiabatic transitions between the initial and the final molecular states induced by nuclear motion. We estimated its probability using the SR {\it ab initio} picture presented above, see Fig.~1(a). In this approximation, it is sufficient to consider only two states: the initial $A$ and the final $X$ ones, both of the
$^1\Sigma^+ (\Lambda=0)$ symmetry, whereas the spin restriction forbids the low-energy CT in the triplet manifold. 

\subsection{Theory}

The nonadiabatic nuclear dynamics was studied within the formalism
of the standard adiabatic (Born-Oppenheimer) approach described, for
example, in Refs.~\cite{MottMassey:1949,MaciasRiera:1982physrep,Grosser:1986, BEGM:2001pra}.
The approach is based on a fundamental simplification, the Born-Oppenheimer
separation of electronic and nuclear motions, leading
to a fixed-nuclei electronic structure calculation and an appropriate
treatment of nuclear motion based on the data calculated in the first step. 
This separation results in the total wave function expanded in terms of products of 
the electronic fixed-nuclei wave functions 
$\Psi_{j}({\bf r},{\bf R})$, the angular nuclear wave functions  
and the radial nuclear wave functions,  $\bf r$ being a set of electronic coordinates. 
The electronic wave functions are calculated as described in the previous section. 
The angular nuclear wave functions are expressed via the generalized spherical harmonics \cite{Grosser:1986, BEGM:2001pra}. 
The radial nuclear wave functions obey the system of
coupled channel equations \cite{Grosser:1986, BEGM:2001pra}. 
In the present case of low-energy Yb$^+$ + Rb collisions, when nonadiabatic transitions occur only between two $^1\Sigma^+$ states, the coupled channel equations for the radial nuclear wave functions 
$\bar{F}^{J,E}_{j}(R)$ are reduced to the following equations (in a.u.) 
\begin{equation}
\begin{array}{lcl}
{\displaystyle \left[-\frac{1}{2\mu}\frac{d^{2}}{dR^{2}}+U_j(R)+\frac{J(J+1)}{2\mu R^{2}}-E\right]\bar{F}^{J,E}_{j}=}\\[7mm]
{\displaystyle \frac{1}{\mu}\sum_{k\ne j}\langle\Psi_j|\frac{\partial}{\partial R}|\Psi_k\rangle\frac{d\bar{F}^{J,E}_{k}}{dR}
+\frac{1}{2\mu}\sum_{k}\langle\Psi_j|\frac{\partial^{2}}{\partial R^{2}}|\Psi_k\rangle \bar{F}^{J,E}_{k}}\\[7mm]
{\displaystyle -\frac{1}{2\mu R^2}\sum_{k}\langle\Psi_j|L_{x}^{2}+L_{y}^{2}|\Psi_k\rangle \bar{F}^{J,E}_{k}\,.}
\end{array}
\label{eq:cce2}
\end{equation}
Here $E$ is the collision energy measured from the asymptotic limit of the initial state $A^1\Sigma^+$, 
$J$ is the total angular momentum quantum number, 
which in case of the $^1\Sigma^+$ states represents simultaneously orbital and rotational momenta of the nuclei,
$\mu$ is the reduced mass of the nuclei, $j$ and $k$ indexes runs over $A$ and $X$, $U_j$ is the adiabatic PEC and three terms in the right-hand side contains first- and second-order radial and angular NACMEs, respectively, defined as the integrals over the electronic coordinates $\bf r$ and calculated as described in the previous Section. 

The coupled channel equations (\ref{eq:cce2}) have their simplest and standard form due to the choice of coordinates: the Jacobi coordinates in which the vector $\bf R$ connects the nuclei (for a fixed-nuclei treatment) and the set of electron coordinates {\bf r} is defined from the center of nuclear mass, the coordinates employed in the {\it ab initio} calculations described above.
In these coordinates a special care, e.g., by means of the reprojection method 
\cite{BEGM:2001pra, GMB:1999pra, Belyaev:2010pra}, 
should be taken in the asymptotic ($R\to\infty$) region for calculating of nonadiabatic transition probabilities 
due to the fact that some radial NACMEs may have non-vanishing values, but in the present case all the treated NACMEs have zero asymptotes, see Fig.~2. 

Due to the large energy splitting and small values of NACMEs, the nonadiabatic transition probabilities are expected to be small at low collision energies, 
and further simplification can be achieved by using the perturbation theory. 
The probabilities for a nonadiabatic transition $A \to X$ due to the radial NACMEs (of both orders) can be approximated as \cite{Belyaev:2007epjd}
\begin{equation}
\begin{split}
&P^\mathrm{RAD}_{AX}(J,E) = {} \\
&\frac{1}{4}\left|\int_{0}^{\infty}\langle\Psi_A|\frac{\partial}{\partial R}|\Psi_X
\rangle\left[F^{J,E}_{A}\frac{dF^{J,E}_{X}}{dR}-F^{J,E}_{X}\frac{dF^{J,E}_{A}}{dR}\right]dR\right|^{2} \,,
                   \label{eq:pertP-rad}
\end{split}
\end{equation}
$F^{J,E}_{X(A)}(R)$ being an unperturbed  scattering radial
wave function (distorted wave) in the channel $X(A)$ normalized by the probability current. 
For transitions induced by the angular NACME of the second order, the equations are similar to the coupled channel equations in a diabatic representation, and the perturbation theory provides the following nonadiabatic transition probability 
\begin{equation}
\begin{split}
&P^\mathrm{ANG}_{AX}(J,E) = {} \\
&\left|\int_{0}^{\infty} \frac{1}{R^2} \left[F^{J,E}_{A}
\langle\Psi_A|L_{x}^{2}+L_{y}^{2}|\Psi_X\rangle F^{J,E}_{X}\right]dR\right|^{2} \,.
                  \label{eq:pertP-ang}
\end{split}
\end{equation}
The nonradiative CT cross sections $\sigma^\mathrm{RAD}_{AX}$ and $\sigma^\mathrm{ANG}_{AX}$ due to the radial and the angular NACME  
are computed as a sum over the total angular momentum quantum number $J$: 
\begin{equation}
\sigma^\mathrm{RAD(ANG)}_{AX}(E)=\frac{\pi g_{i}}{2\, \mu\, E}\,\sum_{J}^{}(2J+1)\,P^\mathrm{RAD(ANG)}_{AX}(J,E)
                          \,,\label{eq:cross}
\end{equation}
$g_{i}$ being a statistical probability for population of the initial channel $i$. 
For the entrance $A^{1}\Sigma^{+}$ channel, $g_{i}=1/4$. 
Because of low collision energies treated the radial wave functions have to be calculated up to large internuclear distances, 
up to several thousands {\AA}ngstroms in the present case.

\subsection{Cross sections}

The calculated $A \to X$ nonradiative cross sections are presented in Fig.~3. 
All calculations were performed for the $^{85}$Rb isotope. The $^{174}$Yb isotope is assumed unless  indicated explicitly. 
It is seen that the CT process induced by the angular NACME is roughly 14 orders of magnitude less efficient than that induced by the radial couplings. The reason is that the total electronic orbital momenta $L$, though not the good quantum numbers, are zero in the asymptotic $R \to \infty$ limit. (In the united ion limit, Bh$^+$, at least the $X$ state should also have $L=0$
\cite{Johnson2002}.) Nonvanishing angular NACMEs can only arise from the minor admixtures of excited states with different angular structure to the adiabatic electronic wave functions. In turn, the cross section due to the radial NACMEs is small on its own: 
the corresponding rate coefficient is roughly of the order of 10$^{-26}$ cm$^3$/s at 1 K temperature.  
This is the consequence of both the large adiabatic splitting and weak interaction of the $A^{1}\Sigma^{+}$ and $X^{1}\Sigma^{+}$ states, 
resulting in small values of the radial NACMEs.

\begin{figure}[ht!]
\includegraphics[angle=0, width=\figwidth]{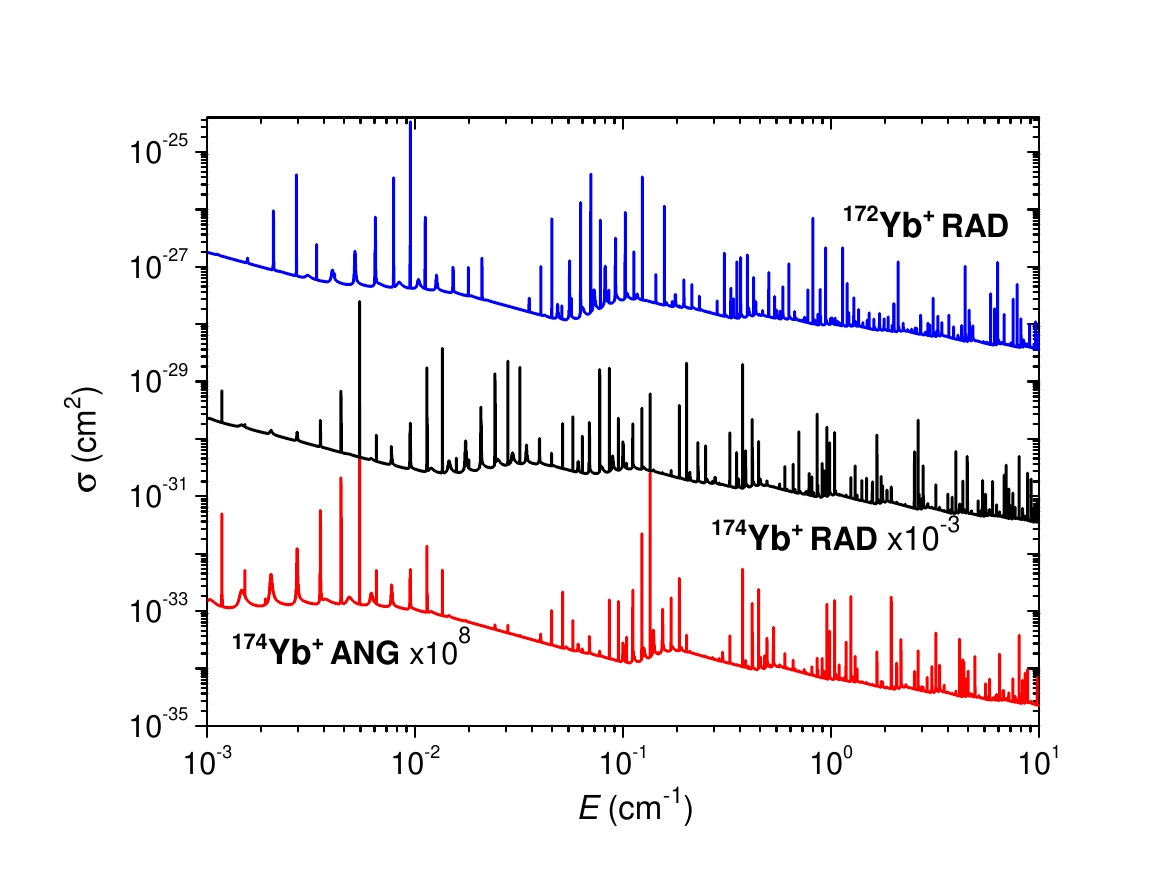}
\caption{\label{fig:nonrad-cs-isotopes}
(color online) 
Nonradiative CT cross sections for Yb$^+$ + Rb collisions induced by NACMEs of different origins. 
The uppermost (blue) trace shows the cross section for $^{172}{\rm Yb}^+$ + Rb collisions due to the radial NACMEs.
The second from the top (black) trace represents the same cross section for $^{174}{\rm Yb}^+$ + Rb collisions (multiplied by $10^{-3}$). 
The bottom (red) trace depicts the cross section for $^{174}{\rm Yb}^+$ + Rb collisions due to the angular NACME (multiplied by $10^{8}$). 
}
\end{figure}

The nonradiative cross sections exhibit numerous resonances whose positions in the radial and angular counterparts are identical for the same Yb isotope. This structure, typical to Langevin regime
\cite{Makarov2003,CoteDalgarno2000,Zhang2011,Zhang2009,Belyaev2011,Belyaev2012}, arises from the tunneling through centrifugal barriers of the effective potential energy in combination with the polarization potential for the initial channel, as is shown in the next Section. Isotope substitution alters the cross section insignificantly. 

To conclude, the large mass of the system, vanishing electronic angular momenta of the $X$ and $A$ states and
their large energy separation in the interaction region suppress the nonadiabatic couplings and make the direct nonradiative CT process very inefficient. 
Inclusion of the SO couplings does not change the conclusion. 
For this reason, we did not exploited more thorough treatments of the nonadiabatic dynamics.

\section{RADIATIVE CHARGE TRANSFER}

The dipole coupling between $A$ and $X$ states should provide more plausible 
radiative CT mechanism. As far as the nonadiabatic effects are negligible, we estimated its efficiency using the standard adiabatic picture and commonly accepted quantum scattering
approaches. Following the previous works on cold ion-atom collisions, we did not consider the effects of external fields and hyperfine structure. 

\subsection{Theory}

Detailed description of the collision-induced spontaneous radiative transition from electronic state $i$ to electronic state $f$
is given by the frequency-resolved cross section $d\sigma^R(E)/d\omega$, where the collision energy $E$ and the final energy $E'$ define the emission frequency $\hbar\omega = E-E'$. 
Depending on the final PEC and the final energy, transitions may occur into the continuum (charge
transfer in the strict sense, hereafter ``charge exchange'', CE) or bound (radiative association, RA) levels. Within the Fermi Golden Rule (FGR)
approximation the corresponding cross sections are expressed as
\cite{Sando1972,Julienne,Julienne-Mies,Zygelman,GGG}
\begin{equation}
\label{diffCT}
\frac{d\sigma^{CE}(E)}{d\omega} = \frac{8\pi^2}{3}\frac{\mu^2}{k^3k'}g_{i}\alpha^3\omega^3
\sum_{J}(2J+1)\sum_{J'}H_{JJ'}D^2_{EJ,E'J'}
\end{equation}
and 
\begin{equation}
\begin{split}
\label{diffRA}
&\frac{d\sigma^{RA}(E)}{d\omega} = \frac{8\pi^2}{3}\frac{\mu}{k^3}g_{i}\alpha^3
\sum_{J}(2J+1) {} \\
&\times\sum_{v'}\sum_{J'}\omega^3_{E,v'J'}H_{JJ'}D^2_{EJ,v'J'}\delta(\omega-\omega_{E,v'J'}).
\end{split}
\end{equation}
Here $J$ and $J'$ are the nuclear angular momenta in the initial and final states, $v'$ is the vibrational 
quantum number of the final state with the energy $E_{v'J'}$, $\hbar\omega_{E,v'J'} = E-E_{v'J'}$, 
$\alpha$ is the fine structure constant, $k^2 = 2\mu E$, $k'^2=2\mu E'$, 
$\delta(x)$ is the Dirac function and $H_{JJ'}$ is the standard H\"onl-London factor. 

The transition dipole moment matrix elements 
on the radial nuclear wave functions 
are defined as
\begin{eqnarray*}
D_{EJ,E'J'} & = & \int_0^\infty F^{J,E}_{i}(R)\, d_{if}(R)\, F^{J',E'}_{f}(R) \, dR, \\
D_{EJ,v'J'} & = & \int_0^\infty F^{J,E}_{i}(R)\, d_{if}(R)\, F^{J',v'}_{f}(R) \, dR.
\end{eqnarray*}
The choice of the coefficients in Eqs.(\ref{diffCT}), (\ref{diffRA}) implies the normalization of distorted waves 
$F^{J,E}_{i}$, $F^{J',E'}_{f}$
used in Ref.~\cite{Zygelman}. 

The total radiative cross section is given by
\begin{eqnarray}
& & \sigma^R_{if}(E)= \sigma^{CE}_{if}(E) + \sigma^{RA}_{if}(E) \nonumber \\
& = & \int_{0}^{\omega_\mathrm{max}}d\omega \left[\frac{d\sigma^{CE}_{if}(E)}{d\omega} + 
\frac{d\sigma^{RA}_{if}(E)}{d\omega}\right] \nonumber \\
& = & \int_{0}^{\omega_0}d\omega \frac{d\sigma^{CE}_{if}(E)}{d\omega} + 
\int_{\omega_0}^{\omega_\mathrm{max}}d\omega \frac{d\sigma^{RA}_{if}(E)}{d\omega},
\label{TotalFGR}
\end{eqnarray}
where $\omega_\mathrm{max}$ is the maximum frequency that corresponds to a transition to the ground level $v'=0$, $J'=0$, 
whereas $\omega_0$ frequency does to the ground-state dissociation limit. 
After substitution of Eq.(\ref{diffRA}) into (\ref{TotalFGR}), integration over $\omega$ with $\delta$-function transforms the last term of the latter into a sum over the discrete $v'$, $J'$ quantum numbers.

The closed expression for the total elastic cross section follows from the optical potential (OP) approach
(see, e.g., Refs.~\cite{Zygelman,Cohen,Tellinghuisen,Gustafsson} and references therein). Combined with
the distorted wave approximation, it reads 
\cite{Zygelman}
\begin{equation}
\label{TotOP}
\sigma^R_{if}(E)  =  \frac{\pi}{k^2}g_{i}\sum_{J}(2J+1)\{1-\exp[-4\eta_J(E)]\}
\end{equation}
with
\begin{equation}
\label{PSrad}
\eta_J(E) = \frac{2\pi}{3}\frac{\mu}{k}\alpha^3\int_0^\infty F^{J,E}_{i}d^2_{if}(R)[U_i-U_f]^3F^{J,E}_{i}dR. 
\end{equation}

For the $A^1\Sigma^+ - X^1\Sigma^+$ or $A0^+ - X0^+$ transitions under consideration, 
$H_{JJ'} = J(2J+1)^{-1}$ if 
$J' = J-1$,  $H_{JJ'} = (J+1)(2J+1)^{-1}$ if $J'=J+1$ and $H_{JJ'}=0$ otherwise. 

\subsection{Cross sections}

The radiative cross sections defined above were computed numerically using fine radial and energy grids and strict convergence criteria for partial wave summations. Accuracy of the OP cross sections was estimated as 3\%, whereas the accuracy of the FGR calculations were 
within 5-8\% due to additional errors in integration over $\omega$ and convergence of the wave functions very close to the dissociation limit of the ground state.

The total OP radiative cross section for the SR model (all the parameters, $A^1\Sigma^+$, $X^1\Sigma^+$ PECs and $d_{AX}$ transition moment function, are taken from the SR calculations) for the collision energy range 0.001 - 10 cm$^{-1}$ are shown in Fig.~4(a) as the upper trace (multiplied by two). It gradually declines with collision energy increase in agreement with the Langevin capture cross section $\sigma_L = 2\pi\sqrt{C_4/E}$ being
five orders of magnitude smaller. The radiative CT cross section exceeds the nonradiative ones by 14 orders of magnitude and bears similar resonance structure. Analysis of the individual partial wave contributions allowed us to correlate the resonance energies with the positions of centrifugal barriers at certain $J$, as is shown in Fig.~5 for the low-energy cross section for collisions involving $^{172}$Yb and $^{174}$Yb isotopes. The cross-section structure consists of  strong narrow resonances, which correspond to the tunneling deep under the centrifugal barriers, and weak broad resonances slightly below or above the barriers. All these findings signify the Langevin character of the Yb$^+$ + Rb cold inelastic collisions.

\begin{figure}[ht!]
\includegraphics[angle=0, width=\figwidth]{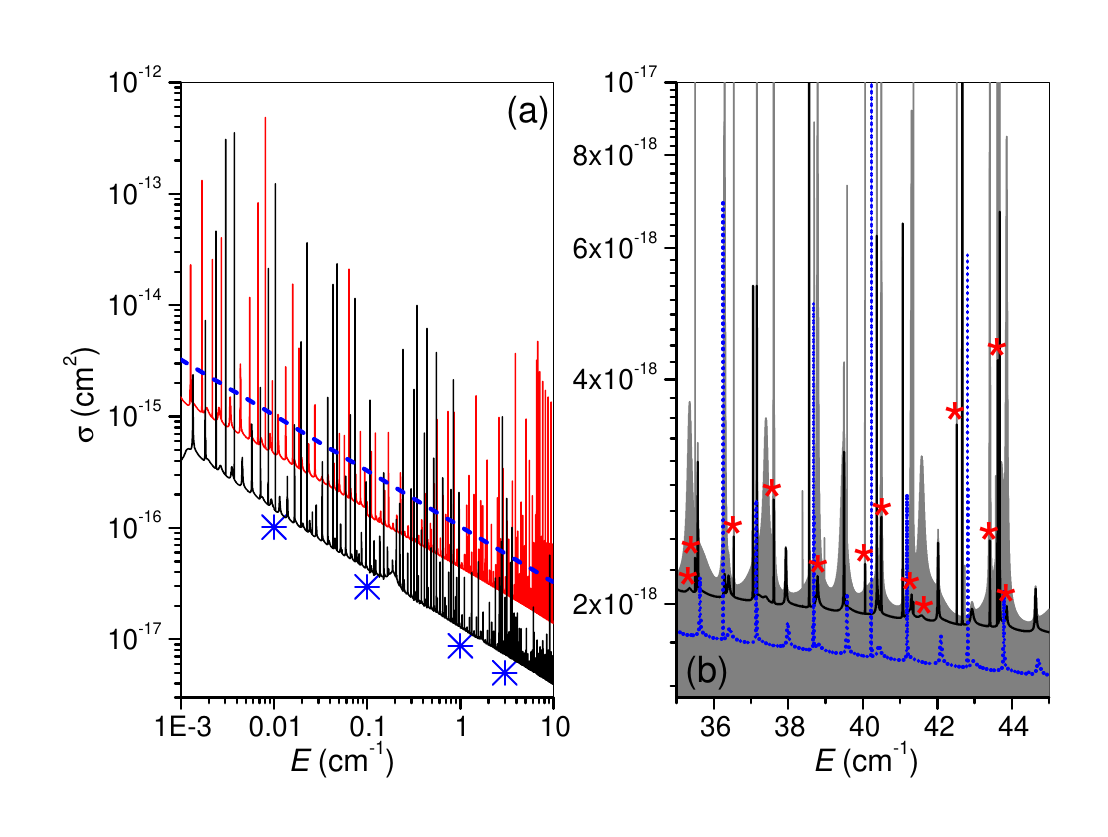}
\caption{\label{fig:radcs}
(color online) Total radiative $A-X$ cross sections for the Yb$^+$ + Rb collisions calculated within the OP approximation. (a) Low-energy results for SR (upper trace, multiplied by two) and SO (lower trace) models. Asterisks represent the RA (free-bound) cross section from the FGR calculations for the SO model. Dashed line corresponds to the Langevin inelastc cross section multiplied by  10$^{-5}$. (b) Cross sections near
the potential barrier top of the $A0^+$ state (40 cm$^{-1}$). Dotted and solid lines -- SR and SO models, respectively. Shaded area presents the results of the ``mixed''  model calculations with the SO-coupled PECs and
SR transition dipole moment. Asterisks indicate the resonance features of the SO cross section associated
with the potential barrier.
}
\end{figure}

\begin{figure}[ht!]
\includegraphics[angle=0, width=\figwidth]{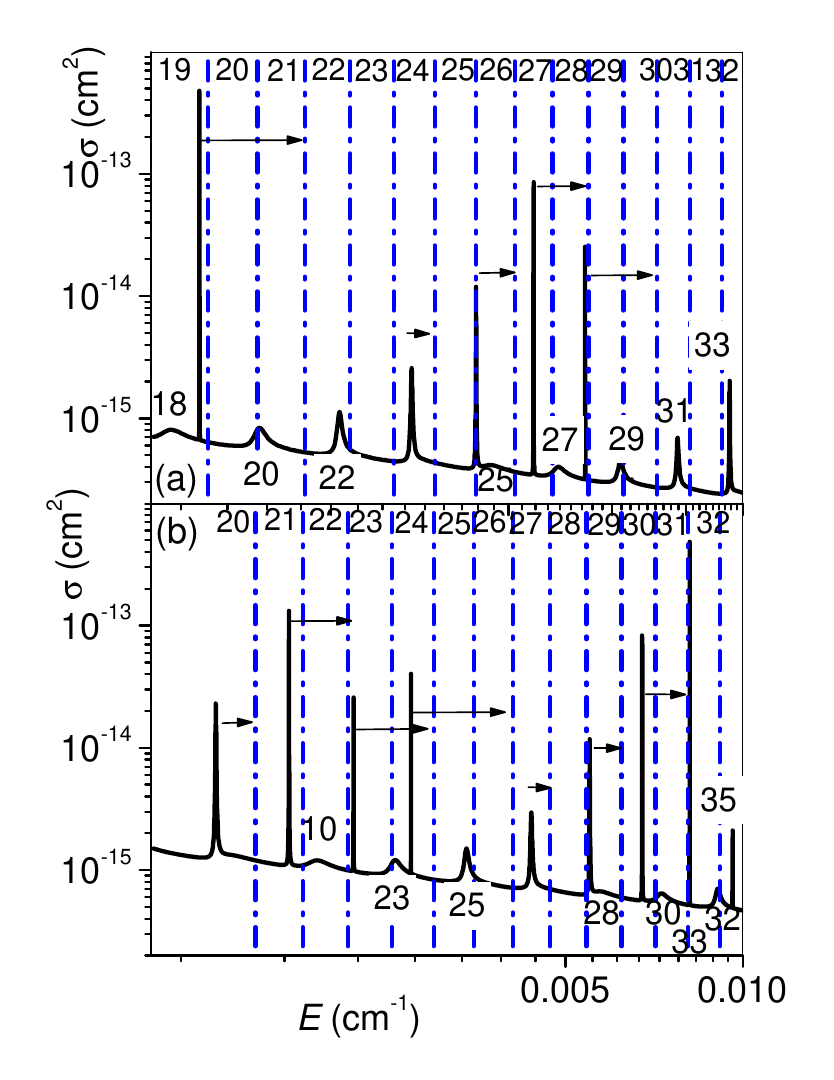}
\caption{\label{fig:radres}
(color online) Low-energy resonance structure of the total radiative $A-X$ cross sections (the SR model, OP
calculations). Vertical lines indicate the positions of centrifugal barrier top at $J$ values given at the top
of each plot. The resonances are assigned to $J$ by arrows or $J$ values given at the bottom of each plot. (a) $^{172}$Yb$^+$ + Rb collisions; (b)  $^{174}$Yb$^+$ + Rb collisions.
}
\end{figure}

The total OP cross section computed within the SO model the SO-coupled $X0^+$, $A0^+$ PECs and the $d_{AX}$ transition moment) are also presented in Fig.~4(a) (lower trace). Test calculations showed that the $a1 -  X0^+$ radiative transition has a negligible probability. The SO model gives the cross section by 10\% larger than the SR one with the same energy dependence, except the ``bumps'' at ca. 0.001 and 0.2 cm$^{-1}$. Resonance structures of two traces are not identical, but the correlations are clearly visible at least below 0.01 cm$^{-1}$ and were confirmed selectively by the partial wave analysis. However, the latter was of little help for seeking the effect of the potential barrier separating two wells of the SO-coupled $A$-state PEC, see Fig.~1(b). 

To elucidate it, we performed OP calculations with the ``mixed'' model that uses SO-coupled PECs but SR transition dipole moment. The results of three models are compared in Fig.~4(b) for collision energies close to the potential barrier top (40 cm$^{-1}$, see Table \ref{potpar}). Preserving the Langevin-type structure similar to that of the SR cross section, the ``mixed'' cross section is dominated by additional very intense features significantly broadened at the background. They were attributed to the potential barrier tunneling or overbarrier resonances, since the penetration into the short-range well enhances the radiative transition probability (both Franck-Condon overlaps and transition moment value are more favorable here than in the long-range well, see Fig.~1). It is not
the case for the consistent SO model, in which the transition moment decline suppresses the radiative transitions from the short-range well. Some features associated to the potential barrier, however, survive in the SO cross sections (most visible are marked by asterisks in the figure). The same ``mixed'' model was used to confirm that the ``bumps'' mentioned above are also due to the potential barrier. As additional comments to Fig.~4(a), increase of the collision energy above the potential barrier does not lead to any drastic change of the cross section because the centrifugal term smoothes it for many partial waves and increase of the SO cross section with respect to SR one reflects the variation of the PECs rather than of the transition moment. 

The results of the FGR calculations performed for the SO model at selected collision energies are presented in Table \ref{crosssec}. The total FGR cross section agrees with the OP one within the numerical accuracy justifying the optical potential approximation. 
Radiative association prevails over charge exchange: within the energy range under study, transitions to the bound rovibrational levels of the  (YbRb)$^+$ ion amount ca. 70\% of the total radiative charge transfer, see Fig.~4(a). Figure 6 compares $A-X$ spontaneous emission spectra at the collision energies 0.01 and 1 cm$^{-1}$. Increase of the collision energy leads mostly to ``rotational'' congestion of the spectrum due to involvement of higher partial waves. It is worthy of noting that all transitions are concentrated in the relatively narrow frequency range, from 500 cm$^{-1}$ above the ground dissociation limit to 2000 cm$^{-1}$ below (ca. 12\% of the available frequency range from 0 to 20250 cm$^{-1}$).

\begin{center}
\begin{table}[ht!]
\caption{
Radiative cross sections (in 10$^{-17}$ cm$^2$) for charge transfer in the Yb$^+$ + Rb
collisions calculated within the SO model  
at several collision energies $E$. Percentage contributions of RA and CE processes are given in parentheses.
}
\label{crosssec}
\begin{threeparttable}
\begin{tabular}{l@{\qquad}c@{\qquad}c@{\qquad}c@{\qquad}c@{\qquad}}
\hline
\hline
$E$, cm$^{-1}$ & \multicolumn{3}{c}{FGR} & OP \\
\cline{2-4}
 &  RA & CE & Total & Total \\
\hline
0.01 & 10.2 (71)& 4.2 (29)& 14.3 & 15.2 \\
0.1 & 3.0 (69)& 1.4 (31)& 4.3 & 4.5 \\
1 & 0.9 (69)& 0.4 (31)& 1.3 & 1.3 \\
3 & 0.05 (67)& 0.03 (33)& 0.08 & 0.08 \\
\hline
\hline
\end{tabular}
\end{threeparttable}
\end{table}
\end{center}

\begin{figure}[ht!]
\includegraphics[angle=0, width=\figwidth]{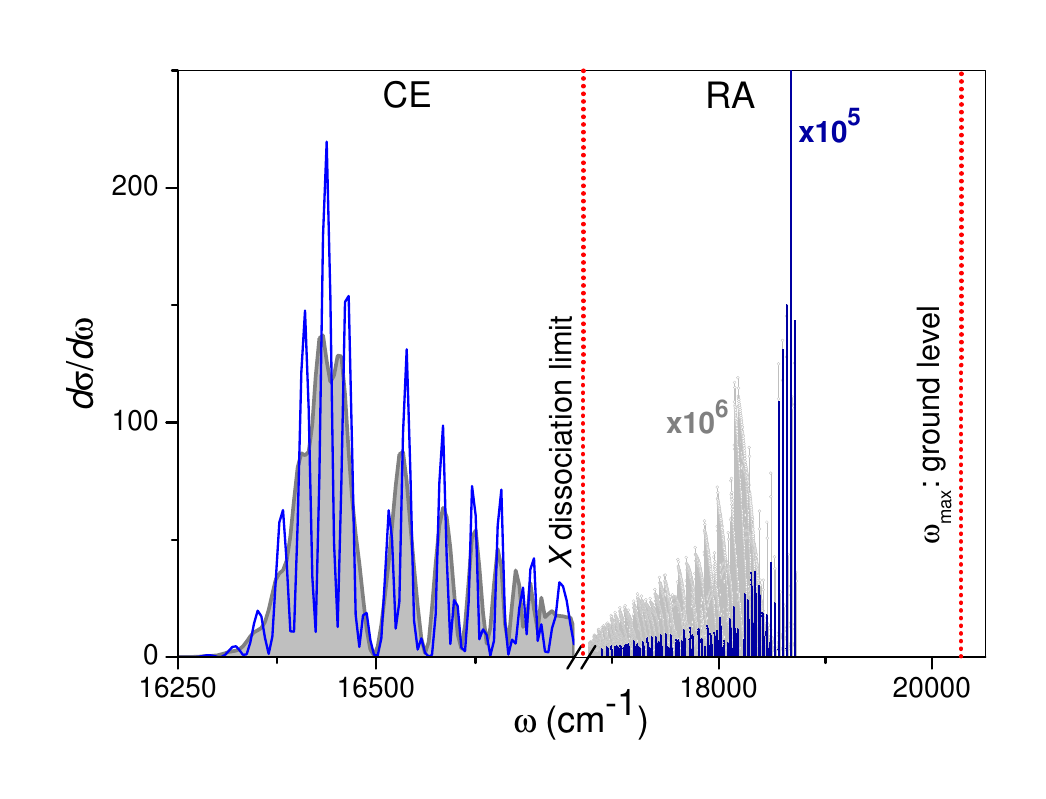}
\caption{\label{fig:spectra}
(color online) 
Frequency-resolved radiative cross sections normalized to the total cross sections, FGR calculations for the SO model. Lines correspond to the
collision energy of 0.01 cm$^{-1}$, shaded areas -- 1 cm$^{-1}$.  Vertical dotted lines mark the borders of free-bound spectrum set by the ground rovibrational level and dissociation limit of the $X$ state. For RA, only the transitions with intensities exceeding 1\% of the maximum one are shown.  
}
\end{figure}

\section{Discussion}

In this Section, we first compare our theoretical results with experimental data available for Yb$^+$ + Rb collisions. Then we discuss them in a broader context of a few previous studies, mainly theoretical, on the cold CT processes in analogous nonresonant ion-atom systems. 

\subsection{Implications to experimental data}

In the experiments by K\"ohl and coworkers 
\cite{Zipkes2010Nature,Zipkes2010PRL} 
the kinetic energy of a single Yb$^+$ ion immersed in an ultracold Rb enesemble was varied by adding excess micromotion energy after displacement of an ion from the center of a trap. The binary-collision ion-loss rate coefficient determined in this way does not correspond to a conventional thermal rate constant for the Maxwell collision energy distribution at certain temperature. For this reason, we also used the effective energy-dependent rate coefficient 
\begin{equation*}
R(E) = \sqrt{2E/\mu}\, \sigma^R_{AX}(E)
\end{equation*}
omitting negligible nonradiative CT contribution. This quantity derived from the OP calculations with the SO model is displayed in Fig.~7 within the energy range probed experimentally 
\cite{Zipkes2010PRL}. 
In agreement with the measurements and the trait of Langevin regime, the background rate coefficient does not depend on a collision energy. To define the average value, we took an arithmetic mean over an uniform grid of 10000 points covering the relevant energy range from 0.15 to 3.25 cm$^{-1}$. For the $^{174}$Yb ion, it amounts to $2.9 \times 10^{-14}$ cm$^3$/s in good agreement with the measured mean of $(4.0\pm 0.3)\times 10^{-14}$ cm$^3$/s
\cite{Zipkes2010PRL}. 
The background is ca. $2.6 \times 10^{-14}$ cm$^3$/s, so the resonances contribute around 10\% to the total rate. 
The SR model gives the mean rate of $R = 2.4\times 10^{-14}$ cm$^3$/s.

\begin{figure}[ht!]
\includegraphics[angle=0, width=\figwidth]{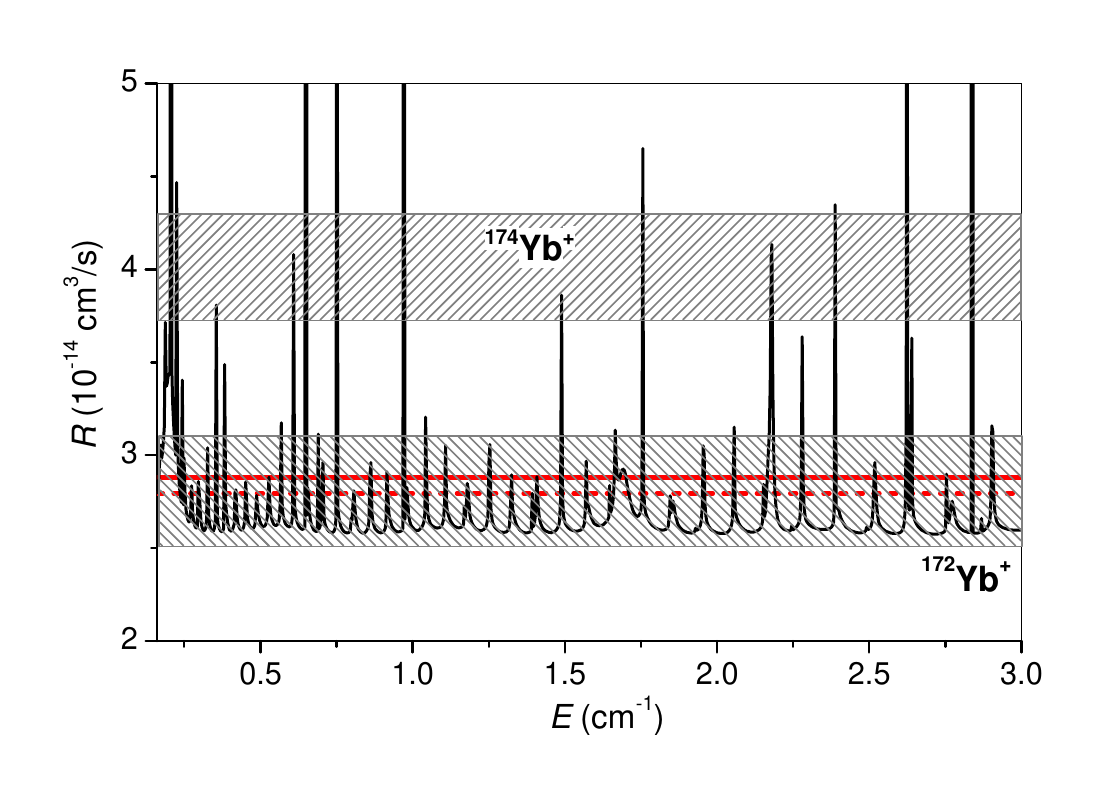}
\caption{\label{fig:RC}
(color online) 
The energy-dependent CT rate coefficient $R_{174}(E)$ (the SO model, OP calculations). Solid and dashed  (red) lines represent the mean values for collisions with the $^{174}$Yb$^+$ and $^{172}$Yb$^+$ ions, respectively. Shaded areas represent the experimental error bars \cite{Zipkes2010PRL}. Rate coefficients are shown in the collision energy range studied experimentally.
}
\end{figure}

Experiments with the $^{172}$Yb$^+$ ions revealed a quite significant isotope effect, $R_{174}/R_{172} \approx 1.4$, see Fig.~7. Using the OP approximation (\ref{TotOP}) and assuming that the terms in the sum do not depend on the reduced mass, one arrives to small reverse scaling, $R_{174}/R_{172} \propto (\mu_{172}/\mu_{174})^{3/2} \approx 0.994$. It is supported by the fact that calculated background rate coefficients for two isotopes are hardly distinguishable. Nevertheless, the mean rate coefficients shown in Fig.~7 reproduce the reverse isotope effect. It originates from the resonant contribution and reflects slightly lower resonance density for the lighter isotope (careful inspection of Fig.~5 indeed reveals the less congested structure). Still, the theoretical rate coefficient ratio 1.03 is much smaller than the measured. More recent experiments for the same system have indicated dramatic effect of the hyperfine structure 
\cite{Ratschbacher2012,KohlPrivate},
which likely contributes to isotope effect as well.

Despite very reasonable agreement for the rate coefficient, our results do not reconcile to the measured product distributions. Experimentally, the probability of Rb$^+$ ion production was found to be 35\%, while the rest 65\% correspond to the loss of charged particle from the trap
\cite{Zipkes2010PRL,Ratschbacher2012}.
Our FGR calculations also give 30\% for CE probability equivalent to Rb$^+$ ion formation, but assign the remaining 70\% to RA, formation of the stable (YbRb)$^+$ molecular ion not observed experimentally. 
Our calculations do not 
identify the trap loss channel, because, according to Fig.~6, all the CE products have the kinetic energy less than 500 cm$^{-1}$ at the trap depth of 1200 cm$^{-1}$
\cite{Zipkes2010PRL}. 
Several works 
\cite{Schmid2010,Rellergert2011,Hall2011,Harter2012} 
emphasized efficient secondary collision processes for Ca$^+$, Ba$^+$ and Rb$^+$ ions in contact with an ultracold Rb ensemble.
This reason, however, was ruled out by Ratschbacher {\it et al.}
\cite{Ratschbacher2012},
who failed to observe (YbRb)$^+$ ion even at very short contact times. We assume that a more plausible explanation can be efficient photodissociation of the molecular ion to highly energetic products by cooling or trapping optical fields. More quantitative estimation may be useful for bridging the gap between theory and experiment. 

\subsection{Other nonresonant cold ion-atom collisions}

In many respects, neutral and singly ionized Yb is the $4f$-analog of an alkaline-earth metal (M). Indeed, a lot of similarity can be traced out in the interactions of Rb atom with Yb$^+$ and M$^+$ ions. Due to mismatch in ionization potentials (IPs), the entrance M$^+$($^2$S) + Rb($^2$S) channel always lies above the ground M($^1$S) + Rb$^+$($^1$S) one, so the location and symmetry of excited M$^*$ + Rb$^+$($^1$S) CT channels determine the main qualitative difference. Binding of $ns^2$ electrons decreases with $n$ from Be to Ba as manifested in IP, $ns^2 \to nsnp$ and $ns^2 \to (n-1)dns$ promotion energies
\cite{NIST}.
By these parameters, Yb most closely resembles Ca. However, in contrast to Yb, the excited  Ca$^*$($^3$P$^\circ$) + Rb CT channel is open and lying slightly below the entrance 
\cite{Hall2011,Belyaev2011}.
Moszynski and coworkers \cite{Krych2011} studied theoretically the Ba$^+$ + Rb system, in which excited CT channels are energetically closed, see also Ref.~\cite{Hall2013arXivBa}.
However, these states correlate to Ba$^*$($^3$D) + Rb$^+$($^1$S) asymptote and exhibit more complicated interaction with the entrance $A^1\Sigma^+$ and $a^3\Sigma^+$ states. In particular, the PEC of the $A$ state has a double-well shape even in the nonrelativistic case. It should be expected that the Sr$^+$ + Rb system reveals more similarity to the Yb$^+$ + Rb one, but we are not aware of any work on it. 

It appears that Yb$^+$ + Rb represents one of the simplest systems from the viewpoint of excited state effects. The use of lighter alkali partners may further simplify the dynamics because increasing IP of the neutral will push the entrance channel down in energy. In the extreme case of Li, it lies only 7000 cm$^{-1}$ above the ground and almost 10000 cm$^{-1}$ below the excited CT asymptote. Collisions with Li should be of special interest for cooling a trapped ion below the Langevin regime, as discussed by Cetina {\it et al.} 
\cite{Cetina2012}. 

Rellergert {\it et al.} \cite{Rellergert2011} investigated cold Yb$^+$ + Ca collisions,
in which only two states of the $^2\Sigma^+$ symmetry separated asymptotically by 1140 cm$^{-1}$ are involved. Avoided crossing at long range presents an interesting feature of this system that facilitate nonadiabatic CT pathway, see below. Similar picture should be expected for Yb$^+$ + Sr, whereas for Yb$^+$ + Ba the CT to excited Ba$^+$ ion becomes possible. Among the nonresonant alkaline-earth systems, Ba$^+$ + Ca  has been studied by Sullivan {\it et al.}
\cite{Sullivan2012}. 
Here the CT process is endothermic and occurs from excited states of the ion. 

It is also instructive to bring together scarce, mostly theoretical, data on the efficiency of distinct CT pathways. To do so on the same footing, we presented $\epsilon_L$ coefficients -- the ratios of the rate coefficient (cross section) to its Langevin inelastic value. 

Our estimation for nonradiative CT rate is vanishingly small, $\epsilon_L \approx 10^{-17}$. Much larger efficiencies, $\epsilon_L \approx 5 \times 10^{-6}$ and 10$^{-3}$, were reported for Yb$^+$ + Ca
\cite{Rellergert2011}
and Ca$^+$ + Rb
\cite{Belyaev2011, Belyaev2012}, 
respectively. In the former case, nonadiabatic transitions should be dramatically enhanced by above mentioned avoided crossing. The latter coefficient in fact refers to the transition to the open excited CT channel, while the efficiency of the direct $A-X$ CT is of order of 10$^{-6}$ 
\cite{Belyaev2011}.
The measured $\epsilon_L \approx 10^{-3}$  for Ca$^+$ + Rb 
\cite{Hall2013arXiv}
confirms the nonradiative CT to excited channels. Indeed, the efficiency of the direct radiative CT was found to be by an order of magnitude lower, $\epsilon_L \approx 7 \times 10^{-5}$ 
\cite{Hall2013arXiv}.  

For radiative CT we got $\epsilon_L \approx 1.5 \times 10^{-5}$, in good correspondence with Ca$^+$ + Rb and  Ba$^+$ + Rb collisions (theory $\epsilon_L \approx 4 \times 10^{-5}$ 
\cite{Krych2011} and $5 \times 10^{-6}$ \cite{Hall2013arXivBa};
experiment 10$^{-4}$ -- 10$^{-3}$  \cite{Schmid2010} and $< 2 \times 10^{-4}$ \cite{Hall2013arXivBa}).
Calculations for Yb$^+$ + Ca provided $\epsilon_L \approx 0.03$, also in agreement with the measured rate
\cite{Rellergert2011}.
The lowest radiative CT efficiency $\epsilon_L \approx 10^{-7}$ was calculated for the Ca$^+$ + Na collisions 
\cite{Makarov2003,Idziaszek2009}.

Even less is known on the product distributions. The molecular ion was observed in the Ca$^+$ + Rb system together with Rb$^+_2$ ion formed in the secondary collisions 
\cite{Hall2011}. 
Studies of the Ba$^+$ + Rb 
\cite{Hall2013arXivBa}
and Yb$^+$ + Ca \cite{Rellergert2011} collisions revealed the same discrepancy between experiment and theory as 
we met here. In the former case, the (BaRb)$^+$ ion constitutes 30\% of the ionic product, while the theory designates it as the barely dominant product of the radiative CT. 
An upper bound for (YbCa)$^+$ formation probability was estimated experimentally as 0.02\%, whereas according to FGR calculations radiative association to molecular ion contributes 50\%. 
It was speculated that the secondary RA processes may deplete the diatomic ion to heavier species like (Ca$_2$Yb)$^+$
\cite{Rellergert2011}.
In our case the measurements failed to detect charged particles different from atomic ions and we proposed alternative explanation -- photodissociation of a molecular ion by cooling and trapping optical field. Finally, calculations on Ca$^+$ + Na collisions showed even stronger predominance of the radiative molecular ion formation: RA to CE branching ratio was found to be around 20:1
\cite{Makarov2003,Idziaszek2009}.

The above overview clearly reveals a variety of the CT mechanisms in cold ion-atom collisions and orders-of-magnitude variations of their rates. The obvious reason is the strong dependence on tiny electronic properties of a system: a relative potential energy gap between initial and final CT channels, transition dipole moments, nonadiabatic couplings. It can be inferred that the radiative pathway generally dominates the direct CT process and leads mainly to the formation of molecular ion. The nonradiative pathway can be competitive if CT occurs to excited states of the products lying close from below to the entrance. The CT processes involving excited ions or atoms 
\cite{Rellergert2011,Hall2011,Ratschbacher2012,Sullivan2012, Belyaev2011}  
may well exhibit a complex interplay of the distinct pathways. 

\section{CONCLUSIONS}

High-level scalar-relativistic {\it ab initio} calculations on the lowest excited electronic states of the (YbRb)$^+$ ion identified the direct
coupling between the entrance $A^1\Sigma^+$ and exit $X^1\Sigma^+$ states as the only pathway for 
charge transfer in the cold Yb$^+$($^2$S) + Rb($^2$S) collisions. Spin-orbit coupling leaves the charge transfer 
through the entrance $a1(^3\Sigma^+)$ state inefficient, but modifies the potential energy curve of the $A$ state to the double-well shape
due to avoided crossing with the excited state correlating to the closed  Yb$^*$($^3$P$^\circ$) + Rb$^+$  
charge-transfer asymptote. 

Quantum scattering calculations showed that the nonradiative charge transfer induced by radial and angular nonadiabatic coupling matrix elements has a 
negligible transition probability. Charge transfer proceeds radiatively with the effective rate coefficient ca. $3 \times 10^{-14}$ cm$^3$/s, that is, 
five orders of magnitude lower than the Langevin capture rate. The calculated radiative cross sections bear all traits of the Langevin 
regime including the rich structure associated with centrifugal barrier tunneling (orbiting) resonances. The short-range well of the 
entrance $A0^+$ state has a weak effect on the charge-transfer efficiency due to decline of the transition moment, but further complicates the resonance structure. 

Calculated radiative charge-transfer rate coefficients agree well with the values of  $(4.0\pm 0.3)\times 10^{-14}$ and $(2.8\pm 0.3)\times 10^{-14}$ cm$^3$/s measured for $^{174}$Yb$^+$ and $^{172}$Yb$^+$ ions, respectively
\cite{Zipkes2010PRL}. 
The observed isotope effect is opposite to that expected from the mass factor for a radiative charge-transfer rate. In our calculations, it is reproduced correctly and originates from the resonance contribution, which increases with the mass of the system. The magnitude of the effect, however, was found to be too small to fully explain the observations. 

The calculations estimated the probability to find the product ion Rb$^+$ in a trap as ca. 30\% in agreement with 35\% obtained experimentally 
\cite{Zipkes2010PRL}. 
The rest of the events correspond to the formation of the bound (YbRb)$^+$ molecular ion, in sharp contrast to observed loss of the charged particle. Efficient molecular ion photodissociation to highly energetic products by cooling and trapping fields may explain this disagreement. 

Present results are in line with the previous studies of similar processes involving alkaline-earth ions and atoms. They suggest that the radiative pathway dominates the ground-state charge transfer having a mean efficiency around 10$^{-5}$ of the Langevin rate and the strong propensity to the molecular ion formation. 

\begin{acknowledgments}
We thank M. K\"ohl, O. Dulieu and T.V. Tscherbul for interest and useful discussions, as well as the referee for fruitful comments. This work was partially supported by the Russian Foundation for Basic Research (projects 11-03-00081 and 13-03-00163) and Russian Academy of 
Sciences (Program of the Fundamental Research by Division of Chemistry and Material Sciences 01 coordinated by Acad. O.M. Nefedov).  
\end{acknowledgments}

\end{document}